\documentstyle[12pt]{article}
\newcommand{\be}{\begin{equation}}
\newcommand{\ee}{\end{equation}}
\newcommand{\cao}{\c c\~ao\ }

\newcommand{\bt} { \begin{tabular} }
\newcommand{\et}{ \end{tabular} }
\newcommand{\bc} { \begin{center} }
\newcommand{\ec}{ \end{center} }

\newcommand{\f}{ \frac }

\newcommand{\la}{\label }
\newcommand{\bfi}{\begin{figure} }
\newcommand{\efi}{\end{figure} }

\newcommand{\btb} { \begin{table} }
\newcommand{\etb}{ \end{table} }
\begin{document}
\title{Conformal invariance studies of the Baxter-Wu model and a related 
site-colouring problem}

\author{F. C. Alcaraz \ \ and \ \  J. C. Xavier \\
	 Departamento de F\'\i sica \\
	Universidade Federal de S\~ao Carlos \\
	 13565-905, S\~ao Carlos, SP, Brasil }
\date{PACS numbers : 05.50+q,64.60Cn, 75.10Jn }
\maketitle
\vspace{0.2cm}
\begin{abstract}
The partition function of the Baxter-Wu model is exactly related to the 
generating function of a site-colouring problem on a hexagonal lattice. We 
extend the original Bethe ansatz solution of these models in order to 
obtain the eigenspectra of their transfer matrices in finite geometries 
and general toroidal boundary conditions. The operator content of these 
models are studied by solving numerically the Bethe-ansatz equations and 
by exploring conformal invariance. Since the  eigenspectra are calculated for 
large lattices, the corrections to finite-size scaling are also calculated.
\end{abstract}
\section{Introduction}
The Baxter-Wu model is defined on a triangular lattice by the
Hamiltonian
\be
H=-J\sum_{<ijk>}\sigma_i\sigma_j\sigma_k ,
\la{ham}
\ee
where the sum extends  over the elementary triangles 
 and $\sigma_i=\pm 1$ are Ising variables located at the sites. 
This model is  self-dual \cite{merlini}
with the same critical temperature as that of  the Ising model on a 
square lattice, and  was
solved exactly in its thermodynamic limit by Baxter and Wu \cite{bw}.
Its leading exponents \cite{bw} $\alpha =2/3$, $\mu=2/3$ and  $\eta=1/4$, are  
 the same as those of the 4-state Potts model \cite{{do},{wu},{baxter}}. 
Due to this and the fact that
both models have a fourfold degenerate ground state, it was
conjectured that they  share the same universality class of critical
behaviour.  However from  numerical studies  of these  models on a
finite lattice it is well known that both models show different
corrections to finite-size scaling. While in the Potts models 
\cite{{cardy},{woyna},{alcbar},{hbb}}
these corrections are governed by a marginal operator, producing
logarithmic corrections with the system size, this is not the case in the
Baxter-Wu model \cite{{hamer},{barber}}. This raises the question of knowing  which 
operator governs these corrections in the Baxter-Wu model.  

With the
developments of  conformal invariance applied to critical
phenomena \cite{cardrev}, two models are considered in  the same universality 
class of critical behaviour only if they have the same operator content, 
not only the leading critical exponents. 
The operator content of the 4-state Potts model was already conjectured
from finite-size studies in its Hamiltonian formulation \cite{baake}, 
and can be obtained by a $Z(2)$ orbifold of the Gaussian model (see 
\cite{ginsparg} for a review).  

In this
letter, by exploiting the conformal invariance at the critical point,  
we report on our calculation of
 the operator content of the Baxter-Wu model. 
In order to do this calculation we have to generalize the original Bethe 
 ansatz solution of the model, since this solution does not give the complete 
eigenspectrum of the associated transfer matrix $T$.
The conformal anomaly $c$ and
anomalous dimensions $(x_1,x_2,...)$ are obtained in a standard way
 from the finite-size behaviour of the eigenspectra of the associated
transfer matrix, at the critical temperature. If we write $T=\exp(-\hat
H)$, then in a strip of width $L$ with periodic boundary conditions the
ground-state energy, $E_0(L)$, of $\hat H$ behaves for large $L$ as
\cite{anomalia}
\be
\frac{E_{0}(L)}{L}= \epsilon_{\infty} - \frac{\pi c v_{s}}{6L^{2}} +o(L^{-2}),
\la{ano}
\ee
where $\epsilon_{\infty}$ is the ground-state energy, per site, in the
bulk limit. Moreover, for  each operator $O_{\alpha}$ with dimension
$x_{\alpha}$ there exists a tower of states in the spectrum of $\hat H$
with eigenenergies given by \cite{{cardrev},{cardim}}
\be 
E_{m,m'}^{\alpha}(L)=E_{0}+\frac{ 2\pi v_s }{ L }(x_{\alpha}+m+m') 
+ o(L^{-1}), \\
\la{dim}
\ee
where $m,m'=0,1,2,\ldots$ . 
The factor $v_{s}$ appearing in (\ref{ano}) and (\ref{dim}) is the
sound velocity and has unit value  for isotropic square lattices. 
The higher
eigenvalues of $T$ can be calculated directly by numerical
diagonalization. However since $T$ is not sparse and has dimension
$2^{L}$, for a horizontal width L,  we cannot compute its eigenspectra
by numerical  diagonalization methods for
$L > L_0 \sim 26$. Instead of a direct calculation we relate this problem to a
{\it{ site-colouring  problem }} ({\bf{SCP}}) 
on a hexagonal lattice, which can be
solved by the Bethe ansatz. Following \cite{bw} the partition function
$Z_{L\times N}^{BW}$, of the Baxter-Wu model on a periodic triangular
lattice with $L$ ($N$) rows  (columns) in the horizontal (vertical) 
direction is
related to the partition function, or generating function,
 $Z_{M\times N}^{SCP}$ of a SCP
 on a hexagonal lattice with $M=2L/3$ ($N$) rows (columns) in the
horizontal (vertical) direction. In the limit $N \rightarrow \infty$, 
\be
Z_{L\times N}^{BW}=Z_{M\times N}^{SCP}.
\la{part}
\ee
In figure 1 the Baxter-Wu model is defined on the triangular
lattice formed by 
connecting the points {\large{$\circ $}}, {\large{$\diamond $}}
   and $\triangle $. 
The related SCP is
defined on the hexagonal lattice  formed by the points $\triangle $ and
 {\large{$\diamond $}}, connected by continuous lines. 
 The configurations in the SCP are defined by attaching at the points of 
the hexagonal lattice 
 $(i,j)$ $(i=1,\ldots, M; j=1,\ldots,N)$ site-colour
 variables $c_{i,j}=1,2,\ldots,8$, satisfying the constraint  that any 
two nearest-neighbour colours must differ by $1$ or $-1$.  
 The partition function  $Z^{SCP}$ given in (\ref{part}) is obtained 
by adding all the colour configurations with weights given by the product 
of the fugacities  
 $z_j(j=1,2,...8)$ of each colour in the lattice configuration. 
These fugacities are 
given by 
\cite{bw}
\be
\begin{array}{l}
z_1=z_3=z_5=z_7= 2\sinh ( 4\beta J) \\
z_2^{-1}=z_4=z_6^{-1}=z_8=\sinh (2\beta J)\equiv t. 
\end{array} 
\label{t}
\ee
The critical point of the Baxter-Wu model and of the SCP is given by 
the self-dual point  $t=t_c=1$. If we write
 $T=\exp(-\hat H)$ for both models and since $Z=Tr(T^N)$, the relation 
(\ref{part}) implies 
\be
Tr(e^{-N H_{L}^{BW}})=Tr(e^{-N H_{M}^{SCP}})
\la{tr}
\ee
Its is important to observe that although $H_{L}^{BW}$ and $H_{M}^{SCP}$
have the same dimension $2^{L}$ they may have different eigenvalues.

In the SCP we say we have a dislocation \cite{bw} in a given link
wherever the colour in its right end is smaller than the left one. The number
$n$ of dislocations in a given row is a conserved quantity.
Consequently the Hilbert space associated to $T^{SCP}$ or $H^{SCP}$ is
separated into block disjoint sectors labelled by the values of $n$. For
a periodic hexagonal lattice of width $M$ (even) the possible values of
$n$ are even and given by $M, M\pm 4, M \pm 8,\ldots , M\pm 4\; int(M/4)$. 
Due to (\ref{tr}) the SCP has also an additional $Z(2)$ symmetry 
(eigenvalues $\epsilon = \pm 1$), 
since adding 4 (modulo 8) to all colours in a given configuration does not change its 
weight in the partition function. 
 The Bethe
ansatz solution presented  by  Baxter and Wu \cite{bw} only gives part of the
eigenspectrum of $H^{SCP}$,  since they considered only eigenstates which are 
even under this symmetry. We generalized \cite{tobepu} their 
solution in order to
obtain the missing odd eigenvectors. For the sake of brevity we only present
here the Bethe ansatz equations. The eigenvalues $E_{n}^{(s_j)}$ of
$H^{SCP}$ in the sector with  $n$ dislocations are given by
\begin{equation}
 E_{n}^{ \{s_i\} } = -\frac{M}{4}\ln (16t^2(1 + t^2)) - 
\sum_{j=1}^n(e_j^{(s_j)}-ik_j^{(s_j)}),
\la{ener}
\end{equation}
where
\be
 e_j^{(s_j)}=1/2\ln\left( x_j+s_j\sqrt{x_j^2-1}\right) ; \; \; 
x_j=\cos(2k_j^
{(s_j)})+t+1/t ,
\la{sinal}
\ee
with $1=s_1=s_2=\ldots=s_{n-l}=-s_{n-l+1}=\ldots=-s_{n}$, and
$l=0,1,...,n$ fixed. The quasi-momenta $\{k_j^{(s_j)}\}$ are obtained by
solving the equations 
\begin{equation}
\exp \left( i M k_j^{(s_j)}\right) =-\sqrt{\epsilon}\prod_{p=1}^n
 \left(\frac{\cosh (e_j^{(s_j)}+ik_p^{(s_p)})}{\cosh (e_p^{(s_p)}+
ik_j^{(s_j)})}\right); \;\; j=1,2,...,n ,
\label{eqb}
\end{equation}
where $\epsilon=\pm 1$. The value $\epsilon =1$ gives the part of the
eigenspectrum which  is even under the $Z(2)$ symmetry and was derived in
\cite{bw}, while $\epsilon=-1$ gives the odd part of the eigenspectrum.
Strictly speaking this is a conjecture since the completeness of the
Bethe ansatz solutions is always a  difficult question.

We have studied extensively these equations numerically at  the critical
point $t=t_c=1$, for general values of $n$, $\epsilon$ and $l$ and for
lattice sizes up to $M \sim 200$. For example, in the eigensector where
$n=M$, by setting $l=0$ we obtain the energies of the 
ground state and first excited state
by choosing $\epsilon =1$ and $\epsilon =-1$, respectively.

Let us calculate the conformal anomaly by using (\ref{ano}). The bulk energy
$\epsilon_{\infty}^{SCP}=-\f{3}{4}\ln 6$ can be obtained from the solution in
the bulk limit \cite{bw} and the sound velocity $v_s=\sqrt{3}/3$, can be
inferred from (\ref{dim}) and an overall analysis of the dimensions appearing in the
model. Using these values in (\ref{ano}) we obtain the finite-size sequence
estimators $c(M)$ for the conformal anomaly, presented in table 1. 
As expected the conformal
anomaly is $c=1$, like the 4-state Potts model. The relation (\ref{tr}) does
not imply that in a finite lattice the eigenergies of $H_{L}^{BW}$ and
$H_{M}^{SCP}$ are the same. In fact this is the case. Fortunately by
comparing their eigenspectra by direct calculations on small lattices we
verify that many of the lower energies, including the ground-state energy, are
exactly the same. Consequently by using the bulk limit value \cite{bw} 
$\epsilon_{\infty}^{BW}=-\f{1}{2}\ln 6$
 we obtain the same sequence shown in table 1. The sound velocity which
comes 
from our analysis is now $v_s=\sqrt{3}/2$ and the conformal anomaly has the
expected value $c=1$.

The large-L behaviour of the energies of excited states will give  the 
operator content of the models. 
Using (\ref{dim}),  the finite-size sequences obtained  for 
some dimensions associated to  $H^{SCP}$ are
shown in table 2. 
The estimators  $x_j^{\epsilon}(M-n,l)$ are  the $j^{th}$  
lowest eigenenergy obtained by solving  (\ref{ener})-(\ref{eqb}) with the 
values  $n,\epsilon$ and $l$.  
 As a result of an
extensive calculation of the eigenspectra of $H^{SCP}$, obtained by
direct diagonalization on small lattices and by solving 
(\ref{ener})-(\ref{eqb}) for large
lattices, we arrive at  the following conjecture. Namely the dimensions of primary 
operators in a given sector 
 labelled by $n=M+4p$  are given by
\be
x_{p,q}=\f{1}{2}(4p^2+\frac {q^2}{4}), \; \; \; q=0,\pm 1, \pm 2,... \; ,
\la{xp}
\ee
where $p = 0, \pm 1, \pm 2, \ldots$ for the periodic lattice.
The number of  descendants, with  dimensions   $x_{p,q}+m+m'$ 
$(m,m' \in {\cal Z})$  
is given by the product of two independent Kac-Moody characters.
 The dimensions (\ref{xp}) are similar, in a Gaussian model \cite{kada} 
with compactification radius equal 2, to
the dimensions of an  operator with vorticity $p$ and  spin-wave number $q$. 
 The Gaussian model at this radius corresponds to the continuum limit 
Kosterlitz-Thouless point of the X-Y model in the torus 
\cite{{kada2},{ginsparg}}. This implies that the SCP belongs to the same 
universality class as the Kosterlitz-Thouless phase transition of the X-Y 
model.

	We have also studied the SCP with more general toroidal boundary 
conditions which preserves the same symmetries as the periodic case (
$n$ and $\epsilon$ are good quantum numbers). 
Those boundary conditions are obtained by imposing  to each  row 
$(j = 1, 2, \ldots)$ of a colour 
configuration in the SCP  the constraint $c_{M+1,j} = c_{1,j}+ k$,
where $k = 0, 2, 4$ or $6$. The periodic case is obtained when $k = 0$. 
The possible values of $n$ are now given by $n = M + 4p$ where 
$p = j - \frac{k}{8}$  ($j = 0, \pm 1, \pm 2, \ldots$ ). 
 In this case $k=2$ and $k=6$ we were not able to apply the Bethe 
ansatz for arbitrary temperatures, but only at the critical temperature 
$t = t_c = 1$ \cite{tobepu}. The Bethe ansatz equations turn out to be 
the same as in (\ref{ener})-(\ref{eqb}) but now the values of $n$ depend on 
the boundary condition. If in (\ref{ano}) we take $E_0(M)$ as the 
ground-state energy of the periodic lattice 
($k=0$) our numerical solutions  of Eqs. 
(\ref{ener})-(\ref{eqb}) also give the dimensions (\ref{xp}) 
but with $p = j - \frac{k}{8}$ ($j = 0, \pm 1, \pm 2, \ldots $). 

Let us return to the Baxter-Wu model. In this case by comparing the
eigenspectra of $H^{BW}$ and $H^{SCP}$, obtained by a direct diagonalization 
on small lattices,  we verify that many of the
dimensions $x_{p,q}$ appearing in (\ref{xp}) are absent. For example the
energies producing the estimators in the $2^{nd}$ and 8th columns
 of table 2 only
appear in $H^{SCP}$. Following for large lattices the energies which
are exactly related in both models we verified that the lower dimensions
in the Baxter-Wu model are given by $x= 0,\frac{1}{8}, \frac{1}{2}, 
\frac{9}{8}, \ldots $, and appear
with degeneracy $d_{x}=1,3,1,9,...$, respectively. These results
supplemented with the global eigenspectrum calculated for small systems,
indicate that the operator content of the Baxter-Wu model is the same as
that of the 4-state Potts model  \cite{baake} and is given in terms of a 
$Z(2)$ orbifold \cite{ginsparg} of the Gaussian model. It is interesting
to remark that while in the SCP the operator content is given in terms
of characters of the Kac-Moody algebra in the Baxter-Wu model the characters
are those of the Virasoro algebra. 
Since the exact integrable SCP and the Baxter-Wu model belong to 
different universality classes some care has to be taken when we import  
exact results from the SCP to the Baxter-Wu model.

We also studied the Baxter-Wu model with more general toroidal 
boundary conditions which preserve its $Z(2) \otimes Z(2)$ symmetry. We 
observe numerically that the eigenenergies which appear in this case can 
also be obtained from the eigenspectrum of the SCP with the toroidal 
boundary conditions we considered ($k = 0,2,4,6$). Calculating the 
corresponding dimensions for large lattices we obtain the same 
dimensions reported in \cite{baake} for the 4-state Potts model. These 
results imply that both models are indeed in the same universality class, 
being governed at the critical point and  arbitrary toroidal 
geometry by the same conformal field theory.

Since we calculate eigenenergies of $H^{BW}$ and $H^{SCP}$ for large
lattices we can now also  
calculate the corrections to finite-size scaling for both
models. Consider the lowest eigenenergy $E_{\alpha}$, associated to an
operator with dimension $x_{\alpha}$. From (\ref{dim}) the correction
$R_{\alpha}(M)$ associated to this level is given by
\be
E_{\alpha}=Me_{\infty}+\f{2\pi v_s}{M}
\left( x_\alpha-\f{c}{12}+R_\alpha (M) \right).
\la{eco}
\ee
According to conformal invariance \cite{cardrev,alcbar} $R_{\alpha}$ should behave as
\be
\la{corre}
R_\alpha (M)=\sum_\gamma \f{a_\gamma}{M^{x_\gamma -2}}+
\sum_{\gamma,\gamma '} \f{a_{\gamma\gamma '}}{M^{x_\gamma + x_{\gamma '} -4}}  ,
\ee
where $\{x_{\gamma }\}$ are the non-relevant  
dimensions $(x_\gamma \geq 2)$ associated to the
operators governing the finite-size corrections. In the 4-state Potts
 model  the lowest dimension $x_{\gamma}$ in this set is 
associated to a marginal operator $(x_{\gamma} = 2)$, 
 and the corrections have a logarithmic behaviour with 
the system size. In table 3 we
show our estimators for 
the dimension $x_\gamma$ of the dominant correction for some 
eigenergies $E_{\alpha}$, with corresponding dimension $x_{\alpha}$, in the 
 SCP and the Baxter-Wu model. In
all these cases we clearly see that $x_\gamma=4$ indicating
that the corrections are integer powers. Rather than the 4-state Potts 
model these corrections  are like those of  the Ising model. This
explains why the finite-size studies of the Baxter-Wu model have good
convergence, in contrast with the 4-state Potts model. 

These
results show that although the 4-state Potts model and the Baxter-Wu model share
the same universality class of critical behaviour, 
having the same operator content, 
 the finite-size effects correspond to
different perturbations of the fixed point of the renormalization group.
The SCP belongs to another universality class. This implies 
that not all exact results derived from the SCP 
should be translated to the Baxter-Wu model.
\begin{center}
{\bf Acknowledgments}
\end{center}

It is a pleasure to acknowledge profitable discussions with M. J. Martins. 
 We also thank M. T. Batchelor for a careful reading of 
our manuscript. This work was supported in part by Conselho Nacional de
Desenvolvimento Cient\'\i fico - CNPq - Brazil, and by Funda\cao de Amparo 
\`a Pesquisa do Estado de S\~ao Paulo - FAPESP - Brazil.
%
%

%
%
%
%%%%%%%%%%%%%%%%%%%
\newpage
\Large 
\bc 
Figure and Table Captions 
\ec
\normalsize
\noindent Figure 1 - The Baxter-Wu model is defined on the triangular 
lattice formed by 
 the points {\large{$\circ $}}, {\large{$\diamond $}}   and $\triangle $. 
The site-colouring problem (SCP)  is
defined on the hexagonal lattice  formed by the points $\triangle $ and
 {\large{$\diamond $}}, connected by continuous  lines.

\vspace{1cm}

\noindent Table 1 - Conformal anomaly estimators $c_M$, as a function of 
$M$, for the SCP and Baxter-Wu model.

\vspace{1cm}

\noindent Table 2 - Scaling dimensions estimators $x_j^{\epsilon}(M-n,l)$, 
as a function of the lattice size $M$, 
for some eigenenergies. These energies are the $j^{th}$ lowest energy 
obtained  by solving 
 (\ref{ener})-(\ref{eqb}) with values $n$, $\epsilon$ and $l$.

\vspace{1cm}

\noindent Table 3 - Estimators of the exponent $x_{\gamma}$ in 
 (\ref{corre}) of the dominant correction of some eigenenergies $E_{\alpha}$ 
with dimension $x_{\alpha}$.

%%%%%%%%%%%%%%%%%%%
%%%%%%%%%%%%%%%%%%%
\newpage
\Large
\bc
Table 1
\ec
\normalsize
%
%%%%%%%%%%%%%%%%%%%%%%
%
\btb[h]
\bc
\bt{||c|c||} \hline
 M    &  $c_M$   \\ \hline\hline
 6    &0.996590995    \\ \hline
 10   &0.998910268      \\ \hline
 50   &0.999959561    \\ \hline
 100  &0.999989915    \\ \hline
 150  &0.999995519     \\ \hline
 200  &0.999997480      \\ \hline
\et
\ec
\etb
%
%%%%%%%%%%%%%%%%%%%
%
\Large
\bc
Table 2
\ec
\normalsize
%
%%%%%%%%%%%%%%%%%%%%%%
%
\btb[h]
\bc
\bt{||c|c|c|c|c|c||} \hline
 M    &  $x_1^{-}(0,0)$ & $x_2^{-}(1,0)$ & $x_3^{+}(0,1)$ & $x_2^{+}(2,0)$ & $x_4^{+}(0,1)$  \\ \hline\hline
 6    &0.12502803 & 0.24896741 & 0.50626226 & 0.62613504 & 0.98648357 \\ \hline
 10   &0.12501702 & 0.24959771 & 0.50215317 & 0.62548322 & 0.99698967 \\ \hline
 50   &0.12500083 & 0.24998323 & 0.50008406 & 0.62502093 & 0.99992101  \\ \hline
 100  &0.12500021 & 0.24999580 & 0.50002099 & 0.62500524 & 0.99998057  \\ \hline
 150  &0.12500009 & 0.24999813 & 0.50000933 & 0.62500233 & 0.99999139  \\ \hline
 200  &0.12500005 & 0.24999895 &  0.50000525 & 0.62500131 & 0.99999516  \\ \hline
\et
\ec
\etb
%
%%%%%%%%%%%%%%%%%%%
%
%
\newpage
\Large
\bc
Table 3
\ec
\normalsize
%
%%%%%%%%%%%%%%%%%%%%%%
%
\btb[h]
\bc
\bt{||c|c|c|c|c|c||} \hline
 M    &  $x_\alpha = 0 $       &  $x_\alpha = 0.125 $    & $x_\alpha = 0.25$ 
        & $x_\alpha = 0.5$  & $x_\alpha = 0.625$    \\ \hline\hline
 20   &4.0836238  & 4.7672008 & 3.953517 & 4.0274781 & 3.9063776 \\ \hline
 50   &4.0466061  & 4.8743240 & 3.990329 & 4.0576819 & 3.9816761  \\ \hline
 100  &4.0035431  & 4.9917793 & 3.998189 & 4.0010866 & 3.9965769  \\ \hline
 150  &4.0012060  & 4.9977042 & 3.999406 & 4.0003372 & 3.9989165  \\ \hline
 200  &4.0005332  & 4.9991458 & 3.999680 & 4.0001660 & 3.9994630  \\ \hline
\et
\ec
\etb
%
%%%%%%%%%%%%%%%%%%%
%

\end{document}